# Molecular Switching Operation in Gate Constricted Interface of $MoS_2$ and hBN Heterostructure

Rahul Tripathi[1], Pritam Bhattacharyya[2], Sukanta Nandi[1], Alok Shukla[2], Abha Misra[1]*

[1]Department of Instrumentation and Applied Physics, Indian Institute of Science, Bangalore, Karnataka, India 560012

[2]Department of Physics, Indian Institute of Technology Bombay, Powai, Mumbai, India 400076

**Abstract**

Combined diverse two-dimensional (2D) materials for semiconductor interfaces are attractive for electrically controllable carrier confinement to enable excellent electrostatic control. We investigated the transport characteristic in heterointerface of multilayer molybdenum disulfide and hexagonal boron nitride ($MoS_2$/h-BN) to reveal that the charge transfer switching (CTS) is highly dependent on both the local gate constriction and bias in the channel. Notably, the CTS is shown to be controlled at a molecular level through electrotunable gated constriction. The resulting significant change in conductance due to exposing 100 parts-per-billion of nitrogen dioxide gas led to a high on/off ratio of $10^2$ for completely switching off the channel thus, acting as a molecular switch. First-principle calculations further explained the mechanism of molecular CTS in the device. The molecular tunability of CTS has not been previously reported in any of the van der Waals semiconductor interfaces. Our finding opens avenues to exploit various atomically thin heterostructures for the mesoscopic transport phenomena towards molecular switching operation at room temperature.

*Email: abha.misra1@gmail.com

**INTRODUCTION**

The electrically modulated switches are one of the most elementary components in electrical circuitry. Molecular-scale switches not only need miniaturization of the devices but also enable multifunctionalities, *e.g.,* robotics, memory, and neuromorphic computing, that are stable, reversible, writable, and addressable in solid-state devices.[1-4] So far, it has been challenging without the chemical process to achieve an efficient molecular switching by any external electrical field to switch the tunnel junction between two stable conductance states in the solid-state device.[5-7] Moreover, the reported molecular switches for memory applications with large on/off ratios ($10^2$–$10^3$) do not provide a non-volatile memory, thus cannot be used in applications that require all-electrically driven nanocircuit components.[2,5-8] Therefore, there is a need for switching elements, which can be useful for the dual-functionality of electrical tunability and high on/off ratio. The *van der Waals* heterostructure (vdWHs) composed of two-dimensional (2D) materials are regarded as promising for the next generation nanodevices because of their unique band structure, enhanced features, and performance.[9,10] The deterministic order of assembly with unprecedented control and accuracy to fabricate complex device architecture constitutes a remarkable paradigm to obtain unusual physical phenomena not observed in the pristine atomic layer.[11,12] Importantly, controlling electron transport with gate-defined constriction allows electrical detection and manipulation of spin, charge, or valley degrees of freedom in these vdWHs.[13-15] Among the family of 2D materials, transition metal dichalcogenides (TMDCs), represented by molybdenum disulfide ($MoS_2$), provide a promising platform due to their distinct electronic and chemical properties.[16-19] In $MoS_2$, the formation of tunneling barriers through nanostructures offers a unique perspective in terms of both carrier-confinement strength and the device geometry, which has been demonstrated recently in many experiments.[15,20,21] Confining charge carriers in $MoS_2$ and forming one-dimensional transport channel by electrostatic gating has been performed by Riccard et al.[21] Lateral stacking of TMDCs offers the electrically controllable carrier confinement, combined with the diverse nature of 2D materials.[14,20-28] The combination of different 2D materials into van der Waals heterostructures creates unprecedented physical phenomena, acting as a powerful tool for future devices. An exhaustive study of the band alignment and relativistic properties of a van der Waals heterostructure is done by Pierucci et al.[29] Furthermore, strong hybridization effects arising between the constitutive layers of a $SnS_2$/$WSe_2$ hetero-bilayer structure was studied by Zirbi et al.[30] The atomic flatness and lack of dangling at the surface of 2D insulators such as hexagonal boron nitride (h-BN) pave the way for a strongly coupled gate-defined quantum devices.[31-33] The ability to manipulate exciton dynamics by creating electrically reconfigurable confinement in $MoS_2$-$WSe_2$ vdWH led to an optical switch at room temperature.[34] Furthermore, such a local electrostatic gating enables a fine-tuning of the motion of electron near the channel with small external perturbation.[15] Layer dependent study of $MoS_2$ transistors for molecular-sensing performances in the presence of gate bias and light irradiation is demonstrated by Late et al.[35] Furthermore, DFT calculations on single-layer and bilayer $MoS_2$ show that the dependence in the charge transfer process in the presence of an applied field. The $MoS_2$ field-effect transistor-based gas sensor has been widely studied because of the field-dependent nature of its electronic states, offering them optimal sensitivity by controlled gating of the sensors.[36-39] The first principle calculation done by Yue et al. demonstrates that all molecules are physisorbed on monolayer $MoS_2$ with small charge transfer, acting as either charge acceptors or donors.[36] The application of a perpendicular electric field can also consistently modify the charge transfer between the adsorbed molecule and the $MoS_2$ substrate. A unique device architecture is designed by Sun et al. to induce tensile strain in bilayer graphene to prevent its mechanical deflection onto the substrate by electrostatic force.[39] Furthermore, the adsorption rate can be controlled by the electric field introduced by applying a back-gate voltage. The significant change in conductance due to quantum confinement is observed earlier in a confined nanostructure in the proximity of an adsorbed molecule.[40,41] In a report by Zhang et al., a large number of gates are employed for the enhanced electrostatic coupling between molecules and the active channel in a single-molecule transistor.[20,42-44] Further, in another report, the combination of single-electron transistors with local gates provides a unique way of charge interaction with an external molecule observed in terms of quantum transport behavior of the system.[45,46] However, electrical manipulation at room temperature is difficult due to the thermal excitation energy ($k_BT$) of the carriers. So far, precise carrier confinement is either enabled through the various arrangement of TMDC layers or controlling the carrier transport by a narrow gating to sense small perturbations. However, far less attention is given to understand the deterministic role of fine-tuned mutual independent gate architecture without applying a narrow pulse. Therefore, independent gates are essential for electrical field-dependent molecular interaction in the presence of external perturbation at room temperature operation.

We report a well-controlled charge transfer switching (CTS) upon molecular interaction in an archetypal vdWH based on a few-layer $MoS_2$ on top of the h-BN layer through electrotunable gated constriction. A thin layer of h-BN is employed on pre-patterned electrodes with variable separations for a fine controlled potential barrier in the channel of $MoS_2$. It enables precise control of electron transport properties upon molecular interaction by independently controlling the bottom gates. The gate combination is essential to separately tune the position of the Fermi level and the carrier density in the channel. Furthermore, the combination of local gates allows defining an electron channel with an off-state when depleted. Transport characteristic revealed that the CTS is a bias dependent conductance phenomenon in the channel due to the gating constriction. The channel conductance can be modulated up to three orders using source-drain bias. Furthermore, the gate combination allows a stepwise change in the conductance across the MoS$_2$ channel, which is observed for a fixed barrier width, and the magnitude is modulated using molecular interaction. Interestingly, the large-conductance change ($10^2$) due to 100 parts-per-billion (ppb) of gas concentration leads to a complete switching off the channel acting as a molecular switch. The electrical field-dependent tunability of CTS has not been previously reported in any atomically thin 2D materials. This observation opens avenues to utilize the variable gate architectural device to investigate mesoscopic transport phenomena for molecular interaction.

**RESULTS**

Devices were prepared by standard mechanical exfoliation of bulk MoS$_2$ and h-BN using the scotch tape technique, as shown schematically in Figure 1(a). A trilayer MoS$_2$ flake (~2 nm thick) is placed on top of a few-layer h-BN (~25 nm thick) flake using the standard dry transfer technique.[47] The bottom h-BN flake is thick enough to ensure an atomically flat surface and separates the MoS$_2$ layer from phonons and charge impurities present in the SiO$_2$ substrate. The flake thicknesses were first determined by the optical contrast and then confirmed by atomic force microscopy, as shown in Figure 1(b). The bottom h-BN works as a dielectric layer for local gates ($V_{local}$), whereas the global gate ($V_{BG}$) is provided with a p-Si/SiO$_2$ (~300 nm) substrate. The device consists of five gate electrodes as shown in Figure 1(c) acting as the local bottom gate with different widths are used to create a variable potential ladder varying from 50 to 1000 nm ($W_{local}$), enabling to evaluate of a systematic study on gate width-dependence on a single device (Figure S1). Figure 1(d) shows the Raman spectra of MoS$_2$ acquired at room temperature with a 532 nm excitation laser. The position of the $E_{2g}^1$ and $A_{1g}$ modes agree well with the available literature on

Raman spectroscopy of thin MoS$_2$ flake. Both modes are sensitive to the number of atomic layers in the MoS$_2$ flake. The separation of $E^1_{2g}$ and $A_{1g}$ peaks were measured to be 22.5 cm$^{-1}$, which confirms the trilayer MoS$_2$ flake.[16] The carrier density in the MoS$_2$ channel is shown to be tuned by biasing the global gate ($V_{BG}$) as in Figure 1(e). Furthermore, a threshold voltage ($V_{th}$) around –12 $V$ is observed with an on/off ratio of 10$^6$, ensuring a well-tuned channel with $V_{BG}$. The carrier concentration in the pristine channel is estimated, $n_{MoS_2} = q^{-1}C_g|V_{th}| = 9.2 \times 10^{11}$ cm$^{-2}$ with $q$ =9.2 × 10$^{-19}$ $C$ and $C_g$ = 1.23 × 10$^{-8}$ $Fcm^{-2}$ for 300 $nm$ thick SiO$_2$ layer.

The output characteristic (source-drain current, $I_{SD}$ versus source-drain voltage, $V_{SD}$) in Figure 2(a) is plotted with different $V_{local}$ for the gate width of $W_{local}$ =1000 $nm$ at room temperature. A dc voltage of 100 mV is applied across the channel results in a linear $I$-$V$ characteristic indicating an Ohmic contact. For, $V_{local}$ = 0 $V$, the modulation of $I_{SD}$ is observed with different $V_{BG}$ from +20 to –20 $V$ in a low bias regime, indicating a typical result for the n-doped device (Figure 2(b)). Furthermore, at positive $V_{BG}$, the conduction in the channel can be monotonically tuned by the local gate (1000 $nm$) at room temperature, down to a completely pinched off state which can be seen in Figure 2(c). The carrier concentration in the channel for the local gate is estimated as $n_{MoS_2}(local) = q^{-1}C_g|V_{th}| = 3.09 \times 10^{11}$ cm$^{-2}$ with $q$ = 9.2 × 10$^{-19}$ $C$ and $C_g$ = 1.41 × 10$^{-7}$ $Fcm^{-2}$ for 25 $nm$ thick h-BN layer.

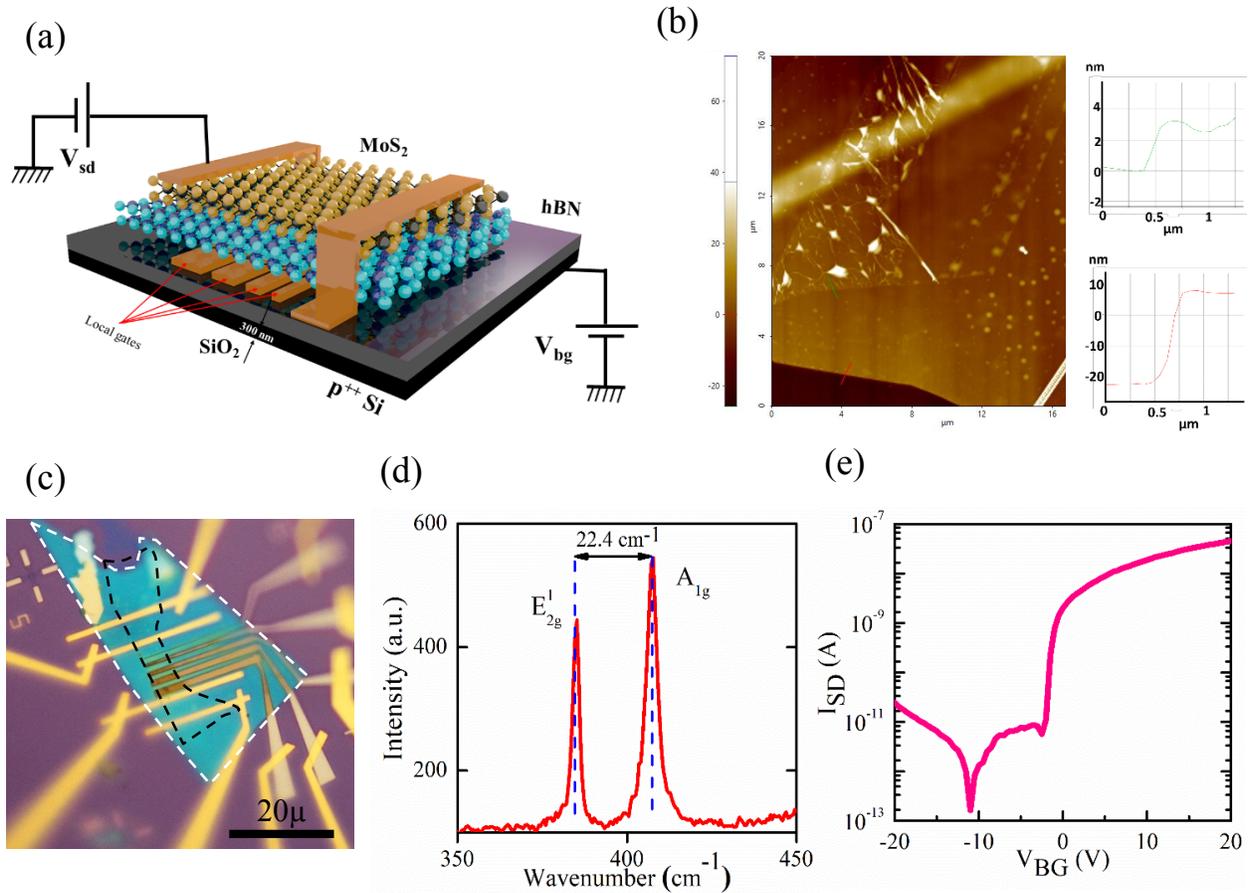

Figure 1. (a) Schematic of the device structure. (b) Microscopic optical image of the device. (c) Atomic force microscopy image of the MoS$_2$ and h-BN flakes. (d) Raman spectra of the MoS$_2$ flake. (e) Conductance measurement of the device with $V_{BG}$.

The variation in drain current using both the $V_{BG}$ and $V_{local}$ shows an interesting transition from ON state to OFF state for a fixed $V_{SD}$. To understand the phenomena, the difference in $I_{SD}$ was carefully analyzed across the ON and OFF states of the device and extracted using $\Delta I_{SD} = (I_{(Vlocal)} - I_{(Vlocal)= -1V})$, as the current change is measured from the offset value ($V_{local}= -1$ V) to observe the change in conductance with switching effect. Furthermore, $\Delta I_{SD}$ is mapped as a function of $V_{BG}$ and $V_{local}$ (for a local gate of 1000 $nm$) (Figure 2(d)). The step-like feature is observed around −2 V of $V_{local}$, indicating a transition from a bulk metallic state to a bulk insulating state within the depleted locally gated region. The flow of the electrons is highly influenced by the direct application of an electric field in the channel, which can be further evaluated using the source-drain bias.

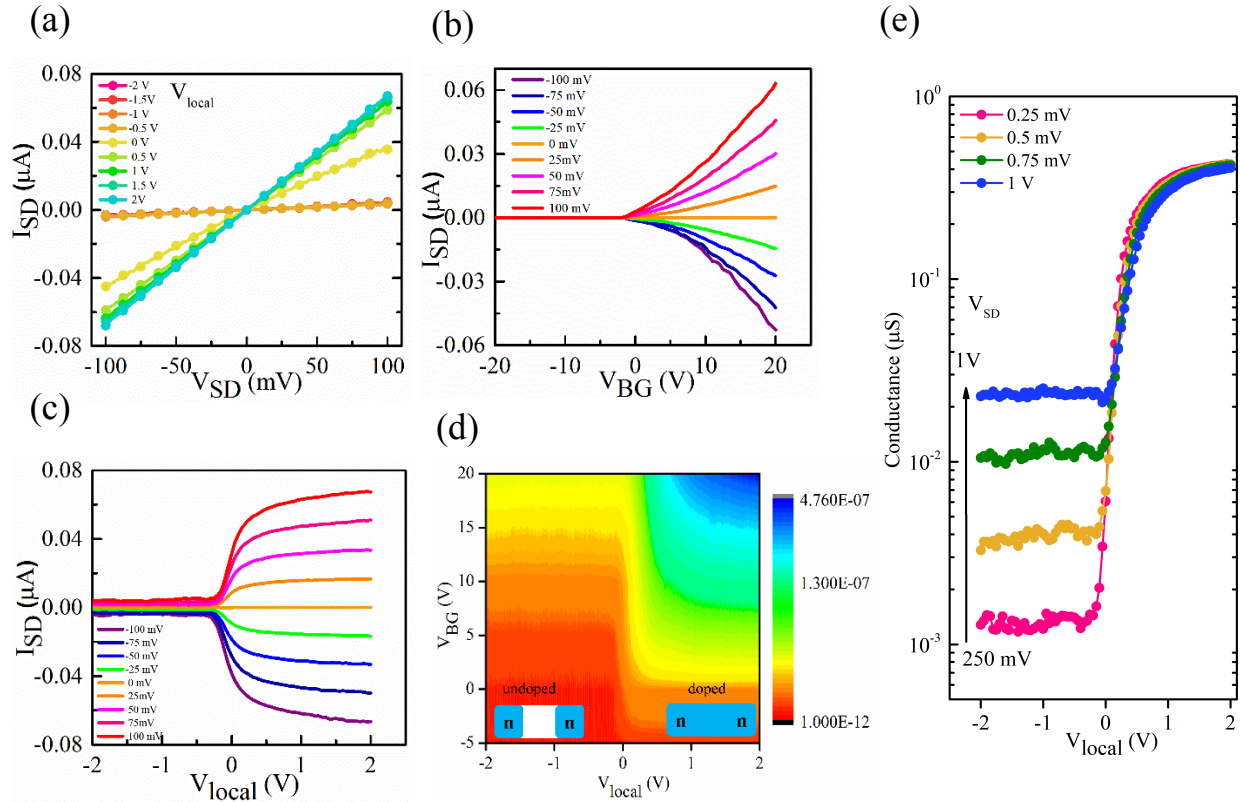

Figure 2. (a) *I-V* measurement of the device. (b) Variation in the drain current with global back-gate voltage. (c) Variation in the drain current with local gate voltage. (d) Color map of the drain current tuned by both $V_{BG}$ and 1000 *nm* wide local gate, $V_{local}$ at room temperature. (e) Conductance plot of the device tuned by different *S-D* bias at fixed $V_{BG}=+20$ *V*, at room temperature.

To eliminate the effect of extra carrier injection due to the applied bias, the conductance (*G*) of the channel is calculated for different $V_{SD}$. Interestingly, a stepwise movement of the conductance is observed in Figure 2(e), which was not observed in the doped region. The conductance of the channel can be tuned monotonically from $10^{-3}$ to $5 \times 10^{-1}$ *μS* by varying $V_{SD}$ from 250 *mV* to *1 V*.

The classical results presented in Figure 2(e) shows an abrupt termination of the charge transfer at the local gate voltage of $V_{local}=-2$ V, indicating the formation of a locally depleted region which is assumed insulating (Figure S2(a)). However, since the region is tunable by electrostatic field, electron tunneling may occur, which would make the channel conductive.[48] Therefore, a quantum mechanical model to account for the electron tunneling is required. Fundamentally, there are two types of tunneling: direct tunneling and Fowler-Nordheim (FN) tunneling.[49] Direct tunneling is significant for small gaps, while FN tunneling dominates if a high electric field is present in the gap. The characteristics of the potential barrier in the channel can further be evaluated with applied bias. Under small applied bias, the direct tunneling current through the barrier can be expressed as

$$I_{SD} \propto V_{SD} \exp\left(-\frac{2d\sqrt{2m^*\varphi}}{\hbar}\right)$$

where $d$, $m^*$ and $\varphi$ are the barrier thickness, effective electron mass and tunneling barrier height, $\hbar$ is the reduced Planck constant. The contribution form direct tunneling current is negligibly small which can be seen in Figure 3(a). At higher source drain voltage, the potential barrier in the channel originates from the vertical electric field and the barrier becomes narrower, consequently leading to the Fowler-Nordheim (FN) tunneling.[50] The FN tunneling current is modeled by

$$I_{SD} \propto V_{SD}^2 \exp\left(-\frac{4d\sqrt{2m^*\varphi^3}}{3\hbar eV}\right)$$

From the above two equations, the plot of $\ln(I/V_{SD}^2)$ versus $1/V_{SD}$ should show a linear regime with a negative slope for the FN tunneling.

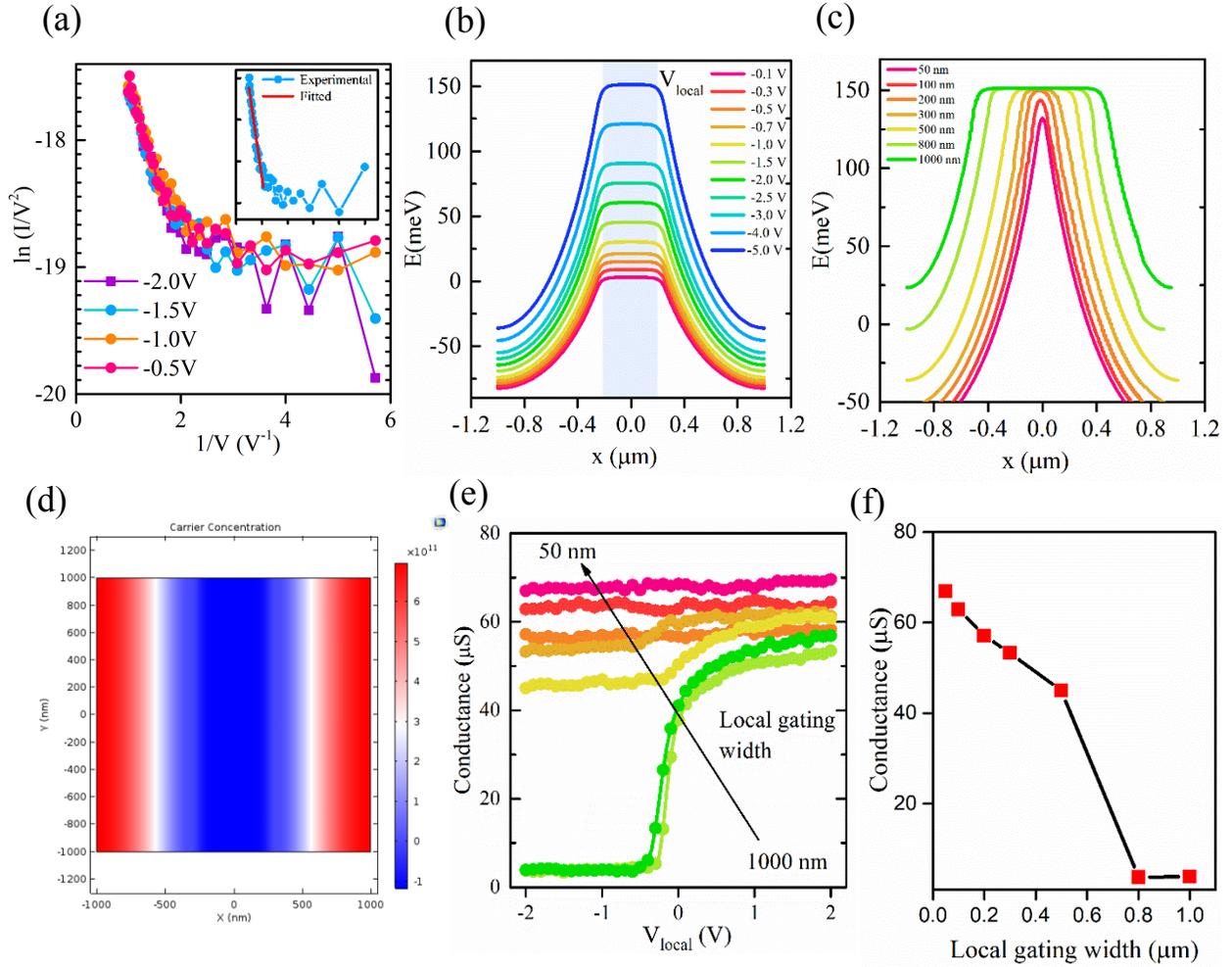

Figure 3. (a) Fowler-Nordheim plots for different local gate voltages for a fixed width of $W_{local}$ =1000 nm. (b) Simulated energy potential profile for different local gate voltages. (c) Simulated potential profile for different width (d) Self-consistent carrier density distribution for local gated geometry at $V_{BG}$ = +4 V and $V_{local}$ = −1.5 V. (e) Conductance plot of the device tuned by different widths of local gate ($V_{local}$) for a fixed S-D bias and $V_{BG}$ = +20 V at room temperature. (f) CTS in the drain current with different widths at a fixed $V_{SD}$ and $V_{local}$ = −1.5 V.

The change in potential barrier height ($\varphi$) with the $V_{local}$ can be extracted from the slope and threshold voltage in Figure 3(a). The threshold voltage for FN tunneling is calculated for all the $V_{local}$ ($V_{local}$ =−2 to −0.5 V) which turns out to be $1/2.49$ to $1/3.67$ V respectively, corresponding to the barrier height ($\varphi/e$) of 0.401 to 0.240 $eV$ in case of 1000 nm local gate width. Inset of Figure 3(a) shows the calculation for $V_{local} = -0.5$ $V$ where barrier is estimated to be about 0.24 $eV$ from the slope, $\frac{4d\sqrt{2m^*\varphi^3}}{3\hbar e} = 1.02$ where $m^* = 0.5m_0$ is the effective mass of the electrons in MoS$_2$. The effective tunneling barrier width is also calculated from the inset of Figure 3(a), which is $d \approx 2.2$ nm. A self-consistent electrostatics simulation has been performed to calculate the potential barrier modulation with $V_{local}$ across the channel. Figure 3(b) shows the modulation of the barrier height with different local gate voltages which is close to the value extracted from the fitting parameters. Interestingly, the vdWHs show a good contrast with conventional devices where the modulation in conductance is mostly obtained in the depletion region. This strongly implies that the vdWHs can be used as the tunneling device with a choice of appropriate local gate bias. The CTS *via* tunneling barrier can be tuned by two different voltages i.e. tunable barrier height by $V_{local}$ and the type of tunneling by $V_{SD}$ (Figure S2 (b, c)). For different widths of local gate, the tunneling barrier energy-space profile is shown in Figure 3(c). In the direct tunneling regime, when the external electric field is small, the tunneling barrier is dominated by the image charge potential. In this regime, the channel conductivity mainly depends on the width as shown in Figure 3(c). Carrier density modulation at the vicinity of the channel is extracted through the electrostatic simulation. The color mapping in Figure 3(d) represents a low charge carrier density near the local gated region at $V_{BG} = +4$ $V$ and $V_{local} = -1.5$ $V$, which is similar to the observed experimental data. To study the effect of the barrier width on the conductance, a series of different widths ($W_{local}$) of local gates were designed in a single channel with the same range of $V_{local}$. Figure 3(e) depicts the conductance as a function of $V_{local}$ for different local gate widths varied from 50 to 1000 *nm*. For the wider channels ($W_{local}$= 1000, 500 *nm*), the conductance is fully controlled from an OFF-state to the ON-state, which can be explained in terms of depletion width. On the other hand, Figure 3(d) depicts there is a stepwise movement in the conductance with lowering in $W_{local}$, which disappeared at $W_{local} = 50$ *nm*, when there is no control over the channel (Figure S3). A temperature dependent study has been performed from 10 to 300 *K* at a fixed bias to observe the effect of temperature on the barrier (Figure S4). When the gap is small, electrons easily tunnel through the narrower barrier, the increased tunneling probability increases the tunneling electron density (channel conductivity).[51,52] The calculated channel conductance is plotted as a function of width for a fixed $V_{local}$ (Figure 3(f)). With the gap of 50 *nm*, the conductance value shows a value of $0.7 \times 10^2$ *μS*, large enough to support the channel conductance even at weak *S-D* electric fields. However, in this direct tunneling regime, the larger gap of 1000 *nm* does not support the charge transfer because the tunneling barrier width is too wide leading to a complete switching off the channel with a low conductance value of ~ $6 \times 10^{-1}$ *μS* which is shown in Figure 3(f).

As discussed earlier, the CTS operation is mainly depending on the width of the local gates and the external electric field on the device, thus shows a potential for the molecular switching operation at room temperature. So far, a switching of the charge carriers is shown in the channel, which is further analyzed under external perturbation, in analogy with the source–drain bias of a conventional vdWHs. The effect of CTS on the molecular sensing ability of the device was evaluated using a controlled nitrogen dioxide ($NO_2$) gas environment. Initially, $NO_2$ gas is injected using mass flow controllers and the concentration of the gas is fixed for 100 ppb throughout the experiment. After 300 s, the injection was stopped. At a regular interval of time, a quick sweeping of local gate voltage was performed ($V_{local}$ = −2 to +2 *V)* at fixed $V_{SD}$ and $V_{BG}$. Because of the nanometer-scale potential barrier width and the inversion layer located near the channel, the molecular interaction is seemingly distinct from other vdWHs. Figure 4(a) shows the response of the device with $W_{local}$ =1000 *nm* in the presence of gas molecules.

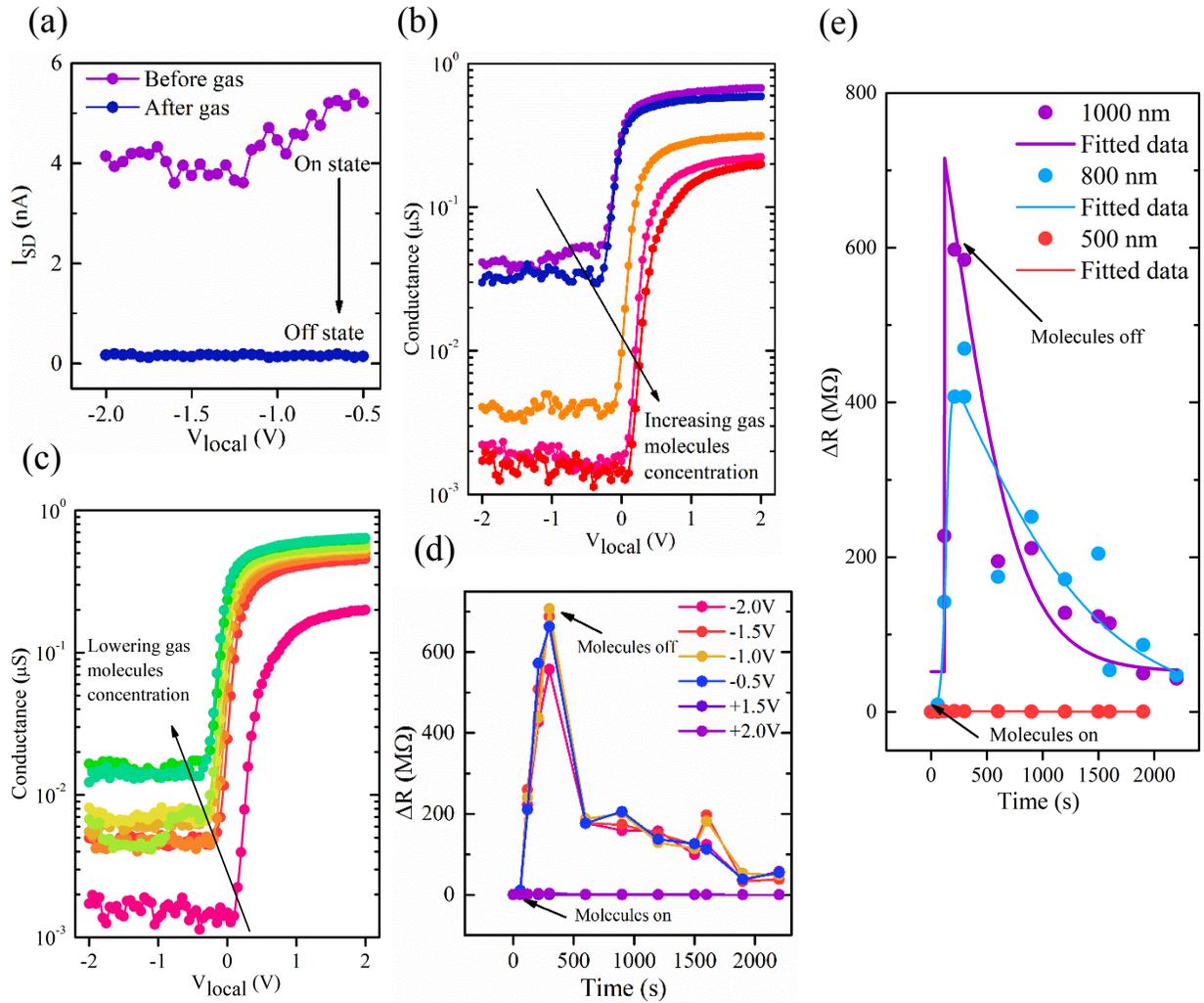

Figure 4. (a) Gate dependence of on/off state with the gas exposure. (b) Systematic change in the conductance with an increment in gas molecule concentration. (c) Switching of the channel from OFF to ON-state by lowering gas molecules concentration. (d) Change in resistance with gas exposure at different local gate voltage. (e) Molecular sensing response for different length of local gates with a non-linear fit of the experimental data.

The initial condition of the device was set to ON-state with a current value of $4.7 \times 10^{-9}$ $A$, which turns to an OFF-state after gas exposure with $I_{SD} = 9 \times 10^{-11}$ $A$. The concentration dependent study is performed by controlling the injection of molecules in the system for 300 $s$. A systematic stepwise decrement in the conductance is observed with the increasing molecular concentration in the channel, which can be seen in Figure 4(b). There is almost a change of $10^2$ in the conductance that occurs with the $V_{local} = -2$ to $-0.5$ V, where the barrier plays a crucial role in the charge conduction. A shift in the $V_{th}$ is observed with the increasing molecular concentration indicating charge transfer mechanism in the vdWHs. Furthermore, the recovery of the device is also studied after the gas injection was stopped. Figure 4(c) depicts a stepwise increment in the conductance with lowering the gas molecule concentration. The change in resistance $\left[\Delta R = (R_{Without\ Gas} - R_{Gas})_{V_{local}=-1V}\right]$ was evaluated in Figure 4(d) with respect to time in presence /absence of molecules for different values of $V_{local}$. Interestingly, a clear dependence of flow of electron on the potential barrier under gas exposure is observed around $V_{local} = -2$ V, which is absent for the $V_{local} = +2$ V. There is a sharp increment in the resistance after the gas exposure that persists up to 300 s and decays slowly to its base value after switching off the gas injection. The peak value of the resistance is observed at around $V_{local} = -1$ V, which vanishes at positive gate bias of $V_{local} = +2$ V. Furthermore, the doping concentration in the channel can be plotted with time which is shown in Figure S5. The value of molecular adsorption rate can be extracted by fitting the curve to an equation $n_d(t) = \frac{n_0 p_a}{a}(exp(-t/a) + n_a)$, where $n_0$ is the number of molecules injected in the system, $n_a$ is the initial adsorption density in the MoS$_2$ and $p_a$ is the adsorption rate. The extracted values of $n_0 p_a$ and $1/a$ are 7.20×10$^{11}$ cm$^2$/min and 0.104%/min, respectively, showing molecules adsorption density of 7.20×10$^{11}$ cm$^2$/min on MoS$_2$ and 0.104% of them desorbing in a minute. Consequently, the Fermi level of the system will further move towards valance band. Furthermore, the tunneling probability of electrons through the barrier will further reduce and the channel will behave as a complete insulator. The length dependence molecular sensing study using a series of local gates of different $W_{local}$ is shown in Figure 4(e). The change in resistance ($\Delta R$) shows the length dependent sensing response with different $W_{local}$ indicating an ON-to-OFF state transition after molecular interaction is more prominent with higher $W_{local}$. Figure 4(e) shows the fitting of the curve using an asymmetric double sigmoid function by an equation $\Delta R = \Delta R_0 + A * \left(\frac{1}{(1+\exp(-(t-B))}\right) * \left(1 - \frac{1}{1+exp(-(t-C))}\right)$, where A, B, C are the constants and $\Delta R_0$ is the initial value of the resistance.

To understand the mechanism behind the CTS via tunneling, the quantum tunneling process is further analyzed under various scenarios. Figure 5 shows schematically how the CTS is critically affected in the presence of molecules which is also verified by density functional theory (DFT) calculations. The energy diagrams corresponding to the specific (hypothetical) case are drawn where the barrier height is shown without the applied bias and with the applied bias at the left contact and at $-2$ $V$ of $V_{local}$ across the channel and no voltage drop at the right contact as shown in Figure 5(a) and 5(b). Enough bias is applied for a transition from a trapezoidal barrier to a triangular barrier as in Figure 5(b) and the position of the Fermi level can be tuned using the fixed $V_{SD}$[53,54] as well as by molecular interaction at a fixed $V_{SD}$. Further, the first-principle calculations were performed using DFT in $NO_2$ adsorbed monolayer $MoS_2$. We considered a 4×4 supercell of monolayer $MoS_2$ with one, two, and four $NO_2$ molecules within the supercell and the band structure of the single $NO_2$ adsorbed system is presented in Figure 5(c).

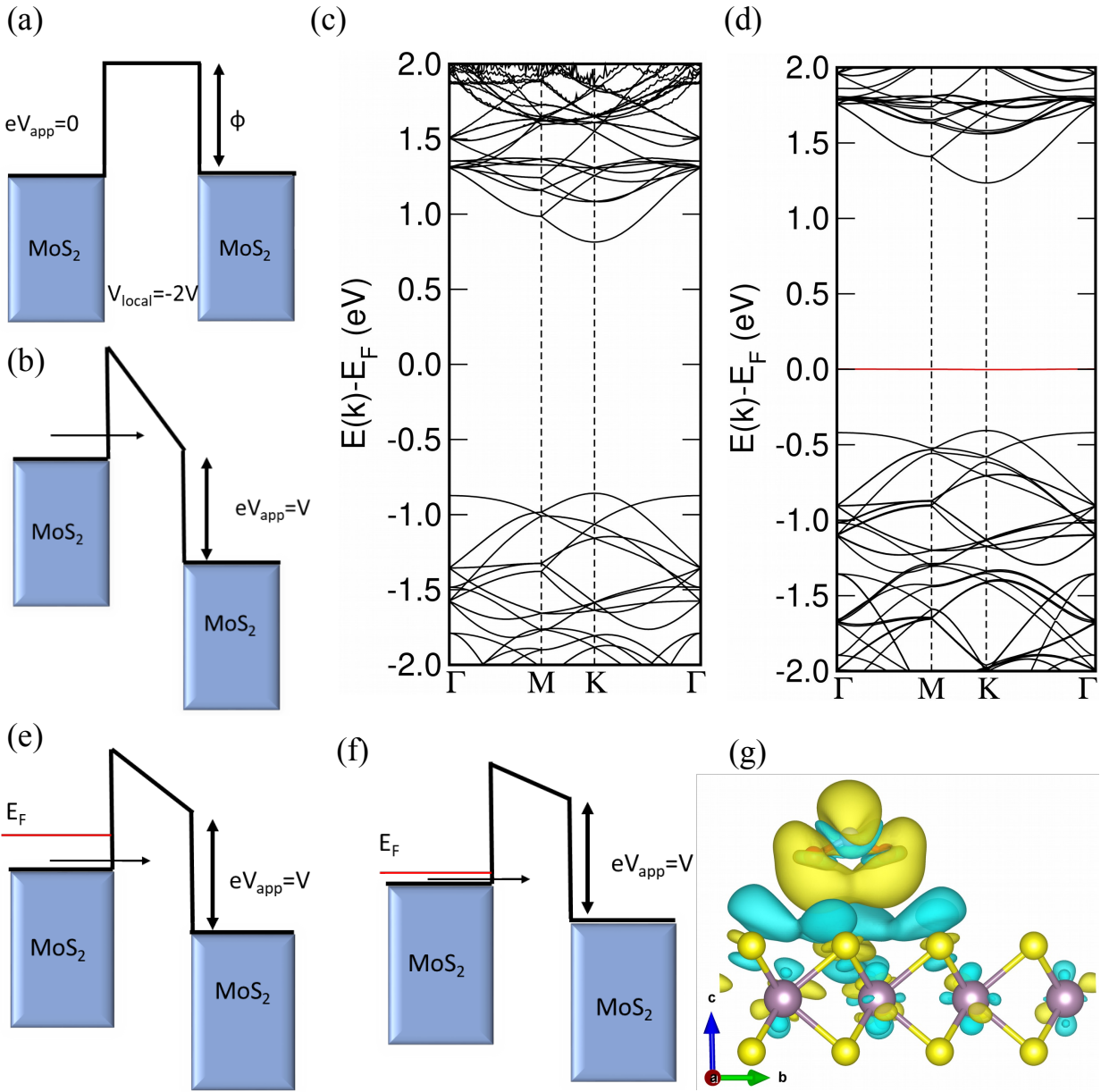

Figure 5. (a) Schematic energy diagram without and (b) with applied source-drain bias ($V_{SD}$). (c) Band structure of pristine $MoS_2$. (d) Change in the Fermi level position for single $NO_2$ molecule adsorbed in a 4×4 supercell of $MoS_2$. (e) Schematic of energy level shift after molecular exposure (f) with increasing molecular concentration. (g) Side-view of the differential charge density plot of the $NO_2$ adsorbed in a 4×4 supercell of monolayer $MoS_2$ system. The electron loss and rich regions are indicated by the cyan and yellow colors.

We observed that the red-colored band near the Fermi level derives predominant contributions from the $NO_2$ molecule. The band structures of the two and four $NO_2$ adsorbed systems are almost similar except firmer shift of the Fermi level towards the valence band. Moreover, the number of energy levels between the valence and the conduction bands increases with the number of $NO_2$ molecules considered in the system. Figure 5(c) shows the band structure of monolayer $MoS_2$, which is in good agreement with the literature.[55] It is obvious from Figure 5(c) and 5(d) that in $NO_2$ adsorbed monolayer $MoS_2$ system, a significant shift of the Fermi level of 0.42 *eV* towards the valence band is observed as compared to the monolayer $MoS_2$.

The shift in Fermi level ($\Delta E_{Fermi}$) with the increasing number of $NO_2$ molecules found out to be 0.50 and 0.57 *eV* for two and four $NO_2$ molecules, respectively. The effect of shift in Fermi level is shown schematically in Figure 5(e) and 5(f) where a transition from a triangular barrier to a trapezoidal barrier with a prominent CTS from ON-state to OFF-state. The differential charge density plots of a single $NO_2$ adsorbed monolayer $MoS_2$ system, depicted in Figure 5(g), confirms the electron transfer from monolayer $MoS_2$ surface to the adsorbed $NO_2$ molecule. The transferred charge are indicated by the cyan and yellow colors, respectively, with an isovalue of $1.98 \times 10^{-4}$ *e/Å³* from vdWHs to gas molecule and the effect of electron transfer, which also indicates the shifting of Fermi level towards the valence band. It is understood that the shift in the Fermi level increases with the concentration of $NO_2$ molecules, which also in accordance with our experimental study.

**CONCLUSION**

A novel approach to control the CTS in depletion region is demonstrated for the first time by engineering the dielectric environment. The multi-gate configuration is used to demonstrate an electrically reconfigurable type of potential landscapes in a single channel. The charge transport through the tunable potential barrier using an in-plane electric field is governed by the carrier tunneling. A strong molecular induced modulation of doping effect is observed in $MoS_2$/h-BN vdWHs and a microscopic description of its origin is also discussed. The carrier movement (block or allow) in the vicinity of the channel with external perturbation is demonstrated, in analogy with source-drain bias of a conventional field effect transistor. Furthermore, the spatial confinement of the charge carriers due the potential barrier leads to an electrically controlled molecular switch using CTS mechanism. The formation of potential landscapes contributes to a new route to design and study quantum phenomena emerging from vdWHs for future study in molecular interactions.

## MATERIALS AND METHODS

### Device fabrication

Heterostructure of $MoS_2$ and graphene was fabricated with two different process for maximum molecular interaction. For the first process a two step lithogrphy process was done on a clean Si/SiO$_2$ wafer without any flake to ensure the flatness and proper transfer of the flakes. In the first step the gate electrode was fabricated followed by deposition proces of Cr/Au (5 nm/10 nm). A standard procedure of dry transfer technique using PDMS (Polydimethylsiloxane) stamp without using any sacrificial layer to ensure the good surface quality. A piece of glass was covered with a PDMS stamp (0.2-0.4 mm thick) which is transparent can be used under the microscope. Few layers of h-BN flake (HQ Graphene) were then exfoliated directly on the stamp and prepared for transfer on top of the local gate electrodes (For details see Supplementary information). Now, $MoS_2$ flake (varying up-to few layers) (Manchester Nanomaterial) was exfoliated onto the PDMS stamp and placed on the micromanipulator stage upside down and aligned perpendicularly with the graphene flake using the microscope. The two were bought into contact and MoS$_2$ flake was transferred on top of h-BN. A 2$^{nd}$ layer of lithography process was done to form the contact pads with a deposition of Cr/Au (10 nm/70 nm).

**Electrical and gas sensing measurements**

Raman spectroscopy was performed in ambient condition, at a wavelength of 532 nm and with a light spot of 1 microns in different places of the device on MoS$_2$-h-BN heterostructure. The electrical characterization of the devices was carried out with an Agilent B1500A semiconductor analyzer in a closed chamber with a nitrogen environment after a stabilization process. Local gate dependent measurement was done by writing a script program for the dual sweep of gate voltage measurement port. Temperature dependent measurement was done in a high-vacuum chamber using Agilent B1500A semiconductor analyzer with a ramp of 25 ºC. All the gas sensing experiments were performed at room temperature and atmospheric pressure. To achieve desirable concentration before being injected into the chamber the nitrogen di oxide gas was mixed with ultra-pure nitrogen gas. The concentration was fixed to 100 ppb of molecular concentration with a total flowrate of 1000 standard cubic centimeters per minute was maintained using the mass flow controllers throughout the experiment.

**Theoretical Approach and Computational Details**

The first principles calculations were performed using density functional theory[56] as implemented in the Vienna Ab-initio Simulation Package (VASP).[57,58] All the calculations were carried out using projector-augmented wave (PAW) pseudo-potentials[57,59] and Perdew-Burke-Ernzerhof (PBE) exchange correlation functional.[60] A kinetic energy cut-off of 500 eV and 5×5×1 k-mesh were considered for structural optimization. The threshold for energy and force convergence was chosen to be $10^{-5}$ eV and $2 \times 10^{-2}$ eV/Å, respectively, while with more than single $NO_2$ molecule in a $4 \times 4$ supercell, the force convergence criteria was set to be $5 \times 10^{-2}$ eV/Å. The density of state calculations was carried out with a tighter threshold for energy convergence, i.e. $10^{-6}$ eV, and using a higher density of k-mesh, i.e. $21 \times 21 \times 1$. At least, a vacuum of 25 Å was considered along the z-direction to minimize the interactions with its images. We also incorporated the DFT-D2 method of Grimme[61,62] to take into account the van der Waals force correction. To perform the band structure calculations, we employed 60 k-points in the reciprocal space. The differential charge density in single $NO_2$ adsorption can be defined as $\rho_{ad} = \rho_{MoS2+NO2} - \rho_{MoS2} - \rho_{NO2}$, where $\rho_{MoS2+NO2}$, $\rho_{MoS2}$ and $\rho_{NO2}$ are the charge densities of the monolayer $MoS_2$ + adsorbed $NO_2$, monolayer $MoS_2$, and isolated $NO_2$ molecule, respectively.

**COMSOL simulation for potential barrier calculation**

To complement and support our analysis of the barrier on the device, we carried out self-consistent electrostatic simulations of the carrier density in the $MoS_2$ plane as a function of the back-gate and local gate voltages. The geometry of the simulation is represented in Fig. S7. The dimensions of the different local gate width and the thicknesses of the $SiO_2$ and h-BN layers are close to those of the measured device. A sigmoid function was considered for simulation of the carrier density in $MoS_2$ sheet (For details see supplementary information). The Poisson's electrostatic static equation was solved with proper boundary conditions for different gate voltages. The profile of the potential energy (E = −eV where e is the charge of the electron and V is the applied voltage at the local gate) across the local gated region is calculated for a fixed back-gate voltage and various local gate voltages and local gate widths.